\documentclass[useAMS,usenatbib]{mn2e}
\usepackage{epsf}
\usepackage{natbib}
\usepackage{epsfig}
\usepackage{subfigure}

\newcommand{\gadget}      {\textsc{gadget-3}}
\newcommand{\gimic}       {\textsc{gimic}}
\newcommand{\subfind}     {\textsc{Subfind}}
\newcommand{\Msun}        {\rm M_\odot}

\title[{Disc galaxies in $\Lambda$CDM}]{Rotation rates, sizes, and star formation efficiencies of a representative population of simulated disc galaxies}

\author[I.~G.~McCarthy et~al.]{I.~G.~McCarthy$^{1,2}$\thanks{E-mail:
mccarthy@star.sr.bham.ac.uk},  J.~Schaye$^{3}$, A.~S.~Font$^1$, T.~Theuns$^{4,5}$, C.~S.~Frenk$^4$, \newauthor R.~A.~Crain$^3$, C.~Dalla Vecchia$^6$
\\
$^{1}$Astrophysics and Space Research Group, School of Physics and Astronomy, University of Birmingham, Edgbaston, Birmingham B15 2TT\\
$^{2}$Kavli Institute for Cosmology, University of Cambridge, Madingley Road, Cambridge, CB3 OHA\\
$^{3}$Leiden Observatory, Leiden University, P. O. Box 9513, 2300 RA Leiden, the Netherlands\\
$^{4}$Institute of Computational Cosmology, Department of Physics, University of Durham, Science Laboratories, South Road, Durham DH1 3LE\\
$^{5}$Department of Physics, University of Antwerp, Campus Groenenborger, Groenenborgerlaan 171, B-2020 Antwerp, Belgium\\
$^{6}$Max Planck Institute for Extraterrestrial Physics, Giessenbachstrabe 1, 85748 Garching, Germany
}

\begin{document}

\date{Accepted ... Received ...}

\pagerange{\pageref{firstpage}--\pageref{lastpage}} \pubyear{2008}

\maketitle

\label{firstpage}

\begin{abstract}
  We examine the rotation rates, sizes, and star formation (SF) efficiencies
  of a representative population of simulated disc galaxies extracted
  from the Galaxies-Intergalactic Medium Interaction Calculation
  (\gimic) suite of cosmological hydrodynamic simulations.  These
  simulations include efficient, but energetically feasible supernova
  feedback, but have not been tuned in any way to produce `realistic'
  disc galaxies. Yet, they generate a large number of discs, without
  requiring extremely high resolution.  Over the wide galaxy stellar
  mass range, $9.0 \la \log_{10}[M_* (\Msun)] < 10.5$, the simulations
  reproduce the observed Tully-Fisher relation, the rotation curves of
  disc galaxies in bins of stellar mass, the mass-size relation of disc galaxies, the optical rotation to virial circular velocity ratio (`$V_{\rm opt}/V_{\rm vir}$'), and the SF efficiencies of
  disc galaxies as inferred from stacked weak lensing and stacked
  satellite kinematics observations.  They also reproduce the specific star formation rates of $\sim L*$ galaxies but predict too low levels of star formation for low-mass galaxies, which is plausibly due to the finite resolution of the simulations.  At higher stellar masses,
  $\log_{10}[M_* (\Msun)] > 10.6$, the simulated galaxies are too
  concentrated and have too high SF efficiencies. We conjecture that
  this shortcoming reflects the neglect of feedback from accreting
  supermassive black holes in these simulations. We conclude that it
  is possible to generate a {\emph {representative}}
  population of disc galaxies that reproduces many of the observed trends of local disc galaxies using standard numerical hydrodynamic
  techniques and a plausible implementation of the ``subgrid''
  astrophysical processes thought to be relevant to galaxy formation.
\end{abstract}


\section{Introduction}

Understanding the formation of normal disc galaxies remains one of the key challenges in astrophysics today.  A general physical picture has been in place for some time (e.g., \citealt{fall80,white91,mo98}), but attempts to form realistic disc galaxies, and to test the general framework in detail, using ab initio cosmological hydrodynamic simulations, have had only limited success thus far.  Specifically, the simulations have historically yielded galaxies with too large of a fraction of their baryons locked up in stars (the ``overcooling problem'', e.g., \citealt{katz96}), sizes that are too small (the ``angular momentum problem'', e.g., \citealt{navarro00}), and rotation curves that decline with radius (rather than being approximately flat as observed) due to the presence of an excessively large spheroidal component (e.g., \citealt{abadi03}).  It is plausible that all three of these problems are related.

It has become increasingly clear from many different lines of evidence
that feedback plays a very important role in regulating the star
formation (SF) of galaxies and also in setting their internal
structure.  Modelling feedback from supernovae (SNe) in simulations of
galaxy formation is challenging, as the injected thermal energy tends
to be radiated away before it can have any hydrodynamical effect
(e.g., \citealt{katz96,balogh01}).  This is likely due to the fact
that simulations still do not have sufficient resolution (nor all the physics necessary) to model the interstellar medium correctly.  Cosmological simulations must
therefore resort to subgrid recipes to try to overcome the overcooling
problem (see the discussions in \citealt{dallavecchia08,dallavecchia12}), which is plausibly the fundamental underlying cause of the other problems
mentioned above (e.g., \citealt{sales10,scannapieco11}).  Typically,
the SN energy is injected kinetically rather than thermally (e.g.,
\citealt{navarro93}) and/or radiative cooling is suppressed by hand
for a short time in the immediate aftermath of energy injection (e.g.,
\citealt{gerritsen97}), although other solutions exist (e.g., \citealt{dallavecchia12}).

In the past few years there has been a good deal of progress on the
formation of more realistic discs galaxies in cosmological
simulations.  A number of groups have been carrying out very high
resolution ``zoomed'' simulations of individual systems primarily
using a thermal SN feedback subgrid prescription (e.g.,
\citealt{okamoto05,okamoto10,governato07,governato10,governato12,stinson10,agertz11,guedes11,avilareese11,maccio12,brook12}; see also \citealt{piontek11} for zoomed simulations with a kinetic feedback model).
Generally speaking, these studies have found that higher 
numerical resolution, together with increased density thresholds for
SF, result in much more efficient feedback which, in turn, allows the
simulations to reproduce many of the properties of normal disc
galaxies, including reasonable SF efficiencies, disc sizes, and
approximately flat rotation curves.  Along with these successes, however, there remain important shortcomings, such as the apparent inability of current simulations (and semi-analytic models) to match the observed specific star formation rates\footnote{The specific star formation rate is defined as the ratio of the star formation rate to stellar mass of a galaxy.} of disc galaxies both locally and at high redshift (e.g., \citealt{avilareese11,dave11}).

While the ability of recent high resolution simulatons to reproduce a number of important observables is certainly encouraging, caution is  
warranted in applying the understanding derived from them to the
galaxy population in general.  Firstly, the feedback implemented in
many of these studies has been tuned to produce `realistic' disc
galaxies.  However, given the uncertain (and possibly degenerate)
nature of subgrid processes, it is dangerous to claim success on the
basis of a handful of simulated galaxies.  The degree of realism
should be judged on the basis of examining many different observables
over a wide range of mass and redshift.  Secondly, the systems that
are chosen to be re-simulated at very high resolution have typically, although not always, been selected on the basis of their merger history - the common lore
being that quiescent merger histories give rise to disc galaxies.
However, the link between merger history and galaxy morphology is
complicated (e.g., \citealt{okamoto05,zavala08,scannapieco09,sales12}) and, in any case, the fact that the majority of isolated normal galaxies in the local Universe are spirals implies that no such selection based on merger history should be necessary.

An alternative approach, therefore, is to try to simulate large,
{\emph {representative}} populations of galaxies without tuning the
feedback or explicitly selecting particular merger histories in order to
produce `realistic' disc galaxies at the present day.  The downside of
this approach is that the resolution must be reduced,
and so too are the kinds of questions one can ask of the simulations.
Also, there is obviously no guarantee that reasonable disc (or
elliptical) galaxies will be produced in the simulation, as there has
been no explicit tuning to get the `right' answer.  At present, it is
computationally feasible to simulate large populations ($\sim 10^3$)
of normal galaxies with $\sim$ kpc scale resolution.  This is the
approach we adopt in the present study. 

In particular, we use the Galaxies-Intergalactic Medium Interaction
Calculation (\gimic) suite of cosmological hydrodynamical simulations,
which were carried out by the Virgo Consortium and are described in
detail by \citet{crain09} (see also \citealt{schaye10}).  The suite
consists of a number of large, nearly spherical regions ($\sim 20
h^{-1}$ Mpc in radius) extracted from the Millennium Simulation and
re-simulated at higher resolution with gas dynamics, including subgrid
prescriptions for metal-dependent radiative cooling, SF,
chemodynamics, and a kinetic implementation of SN feedback.  

The resolution of \gimic\ is somewhat lower than in the studies noted above that have looked at the formation of individual disc galaxies.  Our work is therefore complementary to these very high-resolution studies in that, instead of aiming for maximum resolution on a single system (or a handful of systems), we elect to run lower resolution simulations of a large {\emph {population}} of galaxies, so that we may compare our results to large observational samples of galaxies and place them in a wider context.
We note that a number of recent zoom simulation-based studies have implied that very high resolution is {\emph {required}} in order to form reasonable disc galaxies at all, but the accuracy of this statement depends crucially on the nature of the implementation of important subgrid physics (particularly SN feedback) in the simulations.  As we have shown elsewhere (\citealt{crain10,font11,mccarthy12,sales12}), the \gimic\ simulations have no difficulty in forming disc-dominated (or even bulgeless) galaxies even though they do not have  extremely high resolution (in any case, we show in the Appendix that our results are robust to changes in the resolution).

In the present study, we examine the relation between stellar mass and rotation velocity of disc galaxies (i.e., the `Tully-Fisher' relation), rotation curves in bins of stellar mass of the disc galaxies, the relation between stellar mass and size of disc galaxies, the SF efficiency of both disc and elliptical galaxies, the relation between stellar mass and optical rotation to virial circular velocity ratio ($V_{\rm opt}/V_{\rm vir}$), and the relation between stellar mass and specific star formation rate (sSFR).  We compare the simulations to a variety of recent local observations over a wide galaxy stellar mass range of $9.0 \le \log_{10}[M_* (\Msun)] \le 11.5$.  Throughout we adjust the observational stellar mass measurements so that they are appropriate for a \citet{chabrier03} stellar initial mass function (IMF) (e.g., we add 0.05 dex [subtract 0.1 dex] in the case a Kroupa [`diet Salpeter'] IMF), which is what is adopted in the \gimic\ simulations.  For stellar masses derived using the $M/L$-colour scalings of \citet{bell03} or \citet{bell01}, we also apply a small adjustment to the mass using the expression given in the appendix of \citet{li09} (see also \citealt{dutton11}) so that the stellar masses are consistent with those derived from more accurate `SED fitting' of five-band SDSS data \citep{blanton07}.

The present study is organised as follows.  In Section 2 we present a description of the \gimic\ simulations.  In Section 3 we present our main results on the Tully-Fisher relation, rotation curves, mass-size relation, and SF efficiencies.  In Section 4 we present some scaling arguments that link the faint-end slope of the galaxy stellar mass function to the slope of the Tully-Fisher relation.
 We summarize and discuss our findings in Section 5.

\section{Simulations}

A detailed description of the \gimic\ simulations can be found in \citet{crain09}.  We therefore present only a brief summary here.

The suite consists of a set of hydrodynamical re-simulations of five nearly spherical regions ($\sim 20 h^{-1}$ Mpc in radius) extracted from the (dark matter) Millennium Simulation \citep{springel_etal_05}.  The regions were selected to have overdensities at $z=1.5$ that represent $(+2, +1, 0, -1, -2) \sigma$, where $\sigma$ is the root-mean-square deviation from the mean on this spatial scale.  The 5 spheres are therefore diverse in terms of the large-scale structure that is present within them.  For the purposes of the present study we select `central' galaxies only, which are defined as the most massive self-gravitating collection of stars, gas and dark matter (i.e., the most massive subhalo) in a friends-of-friends group.  Systems and their substructures are identified in the simulations using the \subfind\ algorithm of \citet{dolag09}, that extends the standard implementation of \citet{springel01} by including baryonic particles in the identification of self-bound substructures. 

The cosmological parameters adopted for \gimic\ are the same as those assumed in the Millennium Simulation and correspond to a $\Lambda$CDM model with $\Omega_{m} = 0.25$, $\Omega_{\Lambda} = 0.75$, $\Omega_{b} = 0.045$, $\sigma_{8} = 0.9$ (where $\sigma_{8}$ is the rms amplitude of linearly evolved mass fluctuations on a scale of $8 h^{-1}$ Mpc at $z=0$), $H_{0} = 100 h$ km  s$^{-1}$ Mpc$^{-1}$, $h = 0.73$, $n_s=1$ (where $n_s$ is the spectral index of the primordial power spectrum).  The value of $\sigma_8$ is approximately 2-sigma higher than has been inferred from the most recent CMB data \citep{komatsu11}, which will affect the abundance of the simulated galaxies somewhat, but should not significantly affect their internal properties.

The simulations were evolved from $z=127$ to $z=0$ using the TreePM-SPH code \gadget\ (last described in \citealt{springel05}), which has been modified to incorporate new baryonic physics. Radiative cooling rates for the gas are computed on an element-by-element basis using pre-computed tables generated with CLOUDY \citep{ferland98} that contain cooling rates as a function of density, temperature, and redshift and that account for the presence of the cosmic microwave background and photoionisation from a \citet{haardt01} ionising UV/X-Ray background (see \citealt{wiersma09a}).  This background is switched on at $z=9$ in the simulation, where it `reionises' the simulation volume.  

Star formation is tracked in the simulations following the prescription of \citet{schaye08}.   Gas with densities exceeding the critical density for the onset of the thermo-gravitational instability is expected to be multiphase and to form stars ($n_H \sim 10^{-2}-10^{-1}$ cm$^{-3}$; \citealt{schaye04}). Because the simulations lack the physics to model the cold interstellar gas phase, an effective equation of state (EOS) is imposed with pressure $P \propto \rho^{4/3}$ for densities\footnote{Gas particles are only placed on the EOS if their temperature is below $10^5$ K when they cross the density threshold and if their density exceeds 57.7 times the cosmic mean.  These criteria prevent SF in intracluster gas and in intergalactic gas at very high redshift, respectively \citep{schaye08}.} $n_H > n_*$ where $n_* = 0.1$ cm$^{-3}$.  As described in \citet{schaye08}, gas on the effective EOS is allowed to form stars at a pressure-dependent rate that reproduces the observed Kennicutt-Schmidt law \citep{kennicutt98} by construction.  Note that the resolution of the \gimic\ simulations (see below) was actually chosen so as to marginally resolve the Jeans mass at the adopted density threshold for SF.  Higher resolution is necessary to be able to increase the SF threshold, but it is not sufficient by itself since additional physics is required to accurately capture the phase transition to the cold ISM (certainly radiative transfer but possibly also dust, conduction, cosmic rays, and magnetic fields).  The recent very high resolution zoomed simulations discussed in Section 1, which have pushed to higher SF thresholds, have not included this additional physics. 

The timed release of individual elements by both massive (Type II SNe and stellar winds) and intermediate mass stars (Type Ia SNe and asymptotic giant branch stars) is included following the prescription of \citet{wiersma09b}.  A set of 11 individual elements are followed in these simulations (H, He, C, Ca, N, O, Ne, Mg, S, Si, Fe), which represent all the important species for computing radiative cooling rates, in the presence of a UV/X-ray background.

Feedback from SNe is incorporated using the kinetic wind model of
\citet{dallavecchia08} with the initial wind velocity, $v_w$, set to
$600$ km/s and the mass-loading parameter (i.e., the ratio of the mass
of gas given a velocity kick to that turned into newly formed star
particles), $\eta$, set to $4$.  This corresponds to using
approximately 80\% of the total energy available from SNe for a
Chabrier IMF, which is assumed in the simulation.  Note that unlike many previous works that use similar feedback schemes, the ``kicked'' particles are not hydrodynamically decoupled at any stage.  This leads to efficient entrainment of the ISM by outflows which can significantly affect the properties of galaxies (see \citealt{dallavecchia08}).  The above choice of feedback parameters yields a good match to the peak of the SF
rate history of the universe.  The simulation feedback parameters were
chosen without any regard to observables other than the peak of the cosmic SF history.

To test for numerical convergence, the \gimic\ simulations have been run at three levels of numerical resolution: `low', `intermediate', and `high'.  The low-resolution simulations have the same mass resolution as the Millennium Simulation, while the intermediate- and high-resolution simulations have $8$ and $64$ times better mass resolution, respectively.  Only the $-2\sigma$ and $0\sigma$ volumes were run to $z=0$ at high resolution, owing to the computational expense of these calculations.  We adopt the high-resolution simulations in our main analysis and make comparisons to the intermediate-resolution simulations to test for numerical convergence.  The high-resolution runs have a dark matter particle mass $m_{\rm DM} \simeq 6.62 \times 10^{6} h^{-1}$ M$_{\odot}$ and an initial gas particle mass\footnote{When a star particle forms, it has the same mass as the gas particle from which it formed (the gas particle is removed).  Thus the initial stellar mass is typically $\sim 10^{6} h^{-1}$ M$_{\odot}$. However, star particles lose mass over cosmic time due to stellar evolution.} of $m_{\rm gas} \simeq 1.45 \times 10^{6} h^{-1}$ M$_{\odot}$. The Plummer equivalent gravitational softening is $0.5 h^{-1}$ kpc, which is fixed in physical space for $z \le 3$.

We present a convergence study in the Appendix which shows that the simulation results pertaining to star formation efficiency, kinematics, and sizes are converged for systems with at least $\sim 100$ star particles.  In the analysis below, we adopt a conservative lower stellar mass cut of $\log_{10}[M_* (\Msun)] = 9.0$, which corresponds to approximately $500$ star particles.  We note, however, that a larger number of star particles appears to be required before converged sSFRs are obtained (see the Appendix).

Lastly, the \gimic\ simulations are based on the SPH formalism.  A relevant issue is whether differences in simulation techniques (e.g., SPH, fixed mesh, moving mesh) are important.  It has been shown using idealised tests that standard SPH (which was adopted for the \gimic\ simulations) has a number of problems relating to resolving fluid instabilities and the treatment of weak shocks (e.g., \citealt{agertz07,mitchell09,vogelsberger12,sijacki12}).  While it would be prudent to address these issues in the future (perhaps using modifications to the SPH formalism that appear to mostly resolve these issues; e.g., \citealt{price12,read12}), we point out that both the OWLS \citep{schaye10,sales10} and Aquila \citep{scannapieco11} projects have clearly shown that the differences obtained using different hydrodynamic techniques are small compared to the differences in the results that arise from the use of different subgrid prescriptions.

\section{Results}

Below we examine the $z=0$ stellar mass-rotation velocity
(`Tully-Fisher') relation (Section 3.1), rotation curves in bins of stellar mass (Section 3.2), the mass-size relation (Section 3.3), the SF efficiency of the simulated galaxies (Section 3.4), the stellar mass-$V_{\rm opt}/V_{\rm vir}$ relation (Section 3.5), and the stellar mass-sSFR relation (Section 3.6).  In each case we compare our
results to a variety of recent local observations.  We focus on the
wide stellar mass range, $9.0 \le \log_{10}[M_* (\Msun)] \le 11.5$
(which corresponds roughly to a halo mass range of $11.3 \la
\log_{10}[M_{200}(\Msun)] \la 12.7$ in the simulations, where
$M_{200}$ is the total mass within the radius, $r_{200}$, that
encloses a mean density of 200 times the critical density).  The lower
stellar mass limit is motivated by the numerical convergence study
in the Appendix, while the upper mass limit corresponds to the most
massive galaxies in the simulation volumes.  Within this mass range,
there are 714 simulated galaxies in total in the GIMIC $(-2,0)\sigma$
high-resolution volumes.  This includes both disc- and
spheroid-dominated galaxies.

When making comparisons to the observed Tully-Fisher relation, the mass-size relation, the observed rotation curves of disc galaxies, the $V_{\rm opt}/V_{\rm vir}$, and the sSFRs, it is appropriate to select only disc-dominated systems from the simulations.  One way to do this is kinematically; e.g., to use a cut in the fractional contribution of ordered rotation to the total kinetic energy and/or angular momentum (e.g., \citealt{abadi03,sales10}).  However, in general it is not possible to do this kind of selection for observed galaxies.  Therefore, we instead adopt a simple procedure for assigning galaxies to either the disc- or spheroid-dominated categories based on the distribution of their stellar light.  For each galaxy we produce a projected azimuthally-averaged surface brightness profile\footnote{The i-band luminosities are computed by treating each star particle as a simple stellar population (SSP).  The simulations adopt a Chabrier IMF and store the age and metallicity of the particles.  We use these to compute a spectral energy distribution for each star particle using the GALAXEV model of \citet{bruzual03}.  The i-band luminosity is obtained by integrating the product of the SED with the SDSS i-band transmission filter function.  Our calculation neglects the effects of dust attenuation.} in the SDSS i-band (e.g., as in \citealt{shen03}).  We fit this radial profile with a Sersic law:
\begin{equation}
\mu(R) = \mu_e + \frac{2.5 b_{n}}{\ln(10)}\biggl[(R/R_{e})^{1/n}-1\biggr],
\end{equation}  

\noindent where $R$ is the projected distance from the galaxy center, $R_{e}$ is the effective radius, $\mu_{e}$ is the surface brightness at $R_{e}$, $n$ is the Sersic index, and $b_{n}$ is a function of $n$ only that must be computed numerically (see, e.g., \citealt{graham05}).  Finally, to approximately mimic the depth of SDSS observations, we restrict the fit to radii where $\mu(R) < 24.0$ mags.\ arcsec$^{-2}$.  The results are not sensitive to the exact surface brightness cut adopted.

We use the best-fit Sersic index, $n$, to assign galaxies to the disc-dominated and spheroid-dominated categories.  The case of $n=1$ corresponds to a purely exponential distribution, which is known to describe the disc component of galaxies very well, while $n=4$ corresponds to the classical de Vaucouleurs distribution, which has historically been used to describe the surface brightness profiles of elliptical galaxies.
It has been shown by \citet{blanton03b} using a large sample of galaxies in the SDSS that there is an approximately bimodal distribution in $n$ at fixed galaxy luminosity (or stellar mass), in analogy to the well-known bimodal distribution of galaxy colours at fixed luminosity.  Red galaxies are typically those with large Sersic indices ($n \ga 3$), while blue galaxies correspond to those with relatively small values of $n \la 2$ on average.  Following \citet{shen03}, we use $n=2.5$ as our cut for distinguishing between disc- and spheroid-dominated systems.  With this cut, there are 432 disc-dominated and 282 spheroid-dominated galaxies over the range $9.0 \le \log_{10}[M_* (\Msun)] \le 11.5$ (see Fig.~\ref{fig:fstar} below, which shows the distribution of $n$ with galaxy stellar mass).

\subsection{Stellar mass - rotation velocity relation}

\begin{figure}
\includegraphics[width=\columnwidth]{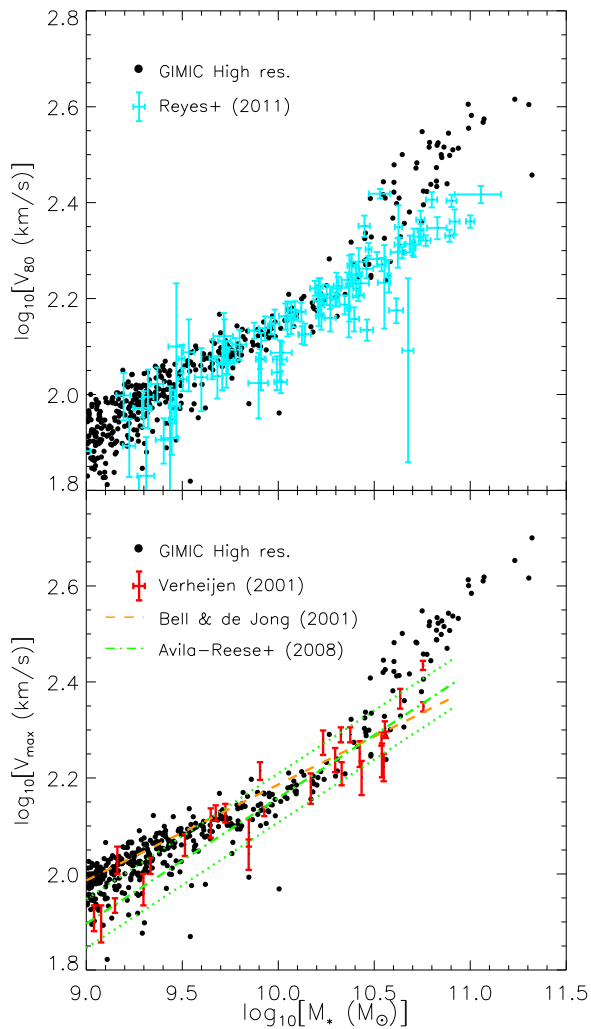}
\caption{\label{fig:tf}
Comparison of simulated and observed relations between galaxy stellar mass and rotation velocity for disc galaxies (i.e., the `Tully-Fisher' relation) at $z=0$.  The top panel shows the case where the velocity is measured at the radius that encloses 80\% of the total i-band flux.  The bottom panel uses the maximum velocity.  Circular velocities are used for the simulated galaxies. The simulated galaxies lie approximately on top of the observed relation over the wide range $9.0 \la \log_{10}[M_* (\Msun)] \la 10.5$.  At higher masses, the simulated galaxies have circular velocities that are too large, which we show later to be due to overcooling in the largest systems.
}
\end{figure}

In Fig.~\ref{fig:tf} we examine the relation between galaxy stellar mass and circular velocity ($V_{c} \equiv \sqrt{G M / r}$) for disc galaxies, which is commonly referred to as the `Tully-Fisher' (hereafter TF) relation.  The top panel shows the relation for the case where the circular velocity is measured at the 3D radius that contains 80\% of the total i-band flux\footnote{We use the best-fit Sersic model to determine the projected radius that encloses 80\% of the flux.} in projection.  We refer to this velocity as $V_{80}$.  The bottom panel shows the relation between the galaxy stellar mass and the maximum circular velocity, $V_{\rm max}$.

We focus first on the top panel of Fig.~\ref{fig:tf}.  The cyan points with error bars correspond to the recent measurements of \citet{reyes11}.  These authors selected a broad sample of galaxies from the SDSS and derived rotation curves from follow-up H$\alpha$ observations (some of which were already presented in \citealt{pizagno07}).  Although \citet{reyes11} did not focus specifically on disc galaxies, the requirement that the galaxies have usable H$\alpha$ rotation curves effectively weeds ellipticals out of their sample.  We use their estimates of the rotation velocity, $V_{80}$, and their published $r$-band magnitudes and $g-r$ colours to convert to stellar mass using the $M/L$ colour scalings of \citet{bell03}.  We select only galaxies from \citet{reyes11} with axis ratios $0.3 < b/a < 0.5$ in order to minimise inclination uncertainties and extinction corrections (as in \citealt{dutton10}).

Over the wide range $9.0 \la \log_{10}[M_* (\Msun)] < 10.5$ the
simulated galaxies fall approximately on top of the observed relation.
To our knowledge, this is the first time a cosmological hydrodynamic
simulation has produced a {\it population} of galaxies that is
consistent with the observed TF relation.  The agreement over this
mass range is not perfect (the observed relation appears to be slightly steeper than the simulated one), but it is clearly a very large improvement over previous simulation studies that looked at representative populations (e.g.,
\citealt{navarro00}).  We note that \citet{derossi10} have also recently examined the TF relation of a population of galaxies simulated with slightly higher resolution but in smaller boxes ($10 h^{-1}$ Mpc on a side).  Their simulated TF relation has approximately the correct slope but no direct comparison was made with observations so it is unclear whether the simulation reproduces the observed zero point as well.

At higher masses ($\log_{10}[M_* (\Msun)] \ga 10.6$), the simulated
galaxies have circular velocities that are significantly larger than the observed rotation velocities. This
is likely the result of overcooling in massive halos (indeed, see
Fig.~\ref{fig:fstar} below).  This may signal that
an increase in the efficiency of SN feedback is required at high
masses and/or that another form of feedback (e.g., from Active
Galactic Nuclei, hereafter AGN) is energetically important at these
mass scales (e.g., \citealt{benson03}).  On the other hand, \citet{derossi10}, who also implemented only SN feedback, did not find such a sharp change in the slope of their TF relation at high masses, although due to their smaller simulation box size they had a much smaller number of high mass galaxies in their sample.  In practice, the mass scale at which overcooling begins to be an issue will depend on the details of the subgrid prescriptions and the adopted parameters, as well as on the resolution.  However, as many studies have demonstrated, it is not obvious that the problems at the massive end can be solved with only feedback from star formation.

It also appears that the simulated TF relation has smaller scatter
than the observed one.  However, it is not clear whether the difference in
scatter is a genuine discrepancy: since we have
measured circular velocities, rather than {\it gas} rotation
velocities (which will not be perfectly circular), the scatter in the
simulated TF relation is likely to be underestimated.  Particle noise
(due to our finite resolution) inhibits the accuracy to which the gas
rotation velocities can be determined in the simulations, which is why
we have used circular velocities instead. 
In addition, scatter in the observed TF relation can be introduced by
errors in corrections for disc inclination and dust attenuation
(although we have attempted to mitigate this by selecting observed
galaxies within a particular axis ratio range), as well as errors in
distance measurements.  A much more careful analysis, taking into
account galaxy selection criteria and observational uncertainties, is
required to make quantitative statements about the degree of
similarity (or lack thereof) in the scatter in the observed and
simulated galaxy populations. 

In the bottom panel of Fig.~\ref{fig:tf} we compare the observed and
simulated TF relations using $V_{\rm max}$ rather than $V_{80}$.  The
red points with error bars represent the 21 cm measurements of
\citet{verheijen01}, who measured the TF relation for a volume-limited
complete sample of spiral galaxies in the nearby Ursa Major Cluster.
We derive stellar masses for their galaxies using their $I$-band
magnitudes and $B-R$ colours and the $M/L$ colour scalings of
\citet{bell01}.  We also show the best-fit power-law relation\footnote{Strictly speaking, this refers to the $V_{\rm flat}-M_{*}$ relation of \citet{bell01}, where $V_{\rm flat}$ is the best-fit asymptotic velocity at large radii.  For galaxies with flat rotation curves $V_{\rm flat}$ is obviously equivalent to $V_{\rm max}$.  For galaxies with rising rotation curves, e.g., low-mass galaxies, $V_{\rm flat} > V_{\rm max}$ as $V_{\rm max}$ is determined from the observed rotation curve whereas $V_{\rm flat}$ represents an extrapolated quantity.} of \citet{bell01} based on K-band data (dashed orange line), as well as that of \citet{avilareese08} based on a compilation from the literature (dot-dashed green line represents their `orthogonal' best fit and dotted green lines represent $\pm 1 \sigma$ intrinsic scatter).  

In accordance with the results shown in the top panel of
Fig.~\ref{fig:tf}, we find that the simulated TF relation (using
$V_{\rm max}$) agrees well with the observed one over the
wide range $9.0 \la \log_{10}[M_* (\Msun)] \la 10.5$ while at higher
masses the simulated galaxies have too high circular velocities with respect to
observed disc galaxies of the same mass. 

In terms of the degree of scatter in the $V_{\rm max}-M_{*}$ relation, we note that the role of distance uncertainties in the observed relation should be minimal for the \citet{verheijen01} results, since their galaxies belong solely to the Ursa Major cluster. Interestingly, the scatter in the $V_{\rm max}-M_{*}$ relation of \citet{verheijen01} and \citet{avilareese08} appears to be similar to that of the simulations for galaxies with $\log_{10}[M_* (\Msun)] \la 10.0$.

\begin{figure*}
\includegraphics[width=15cm]{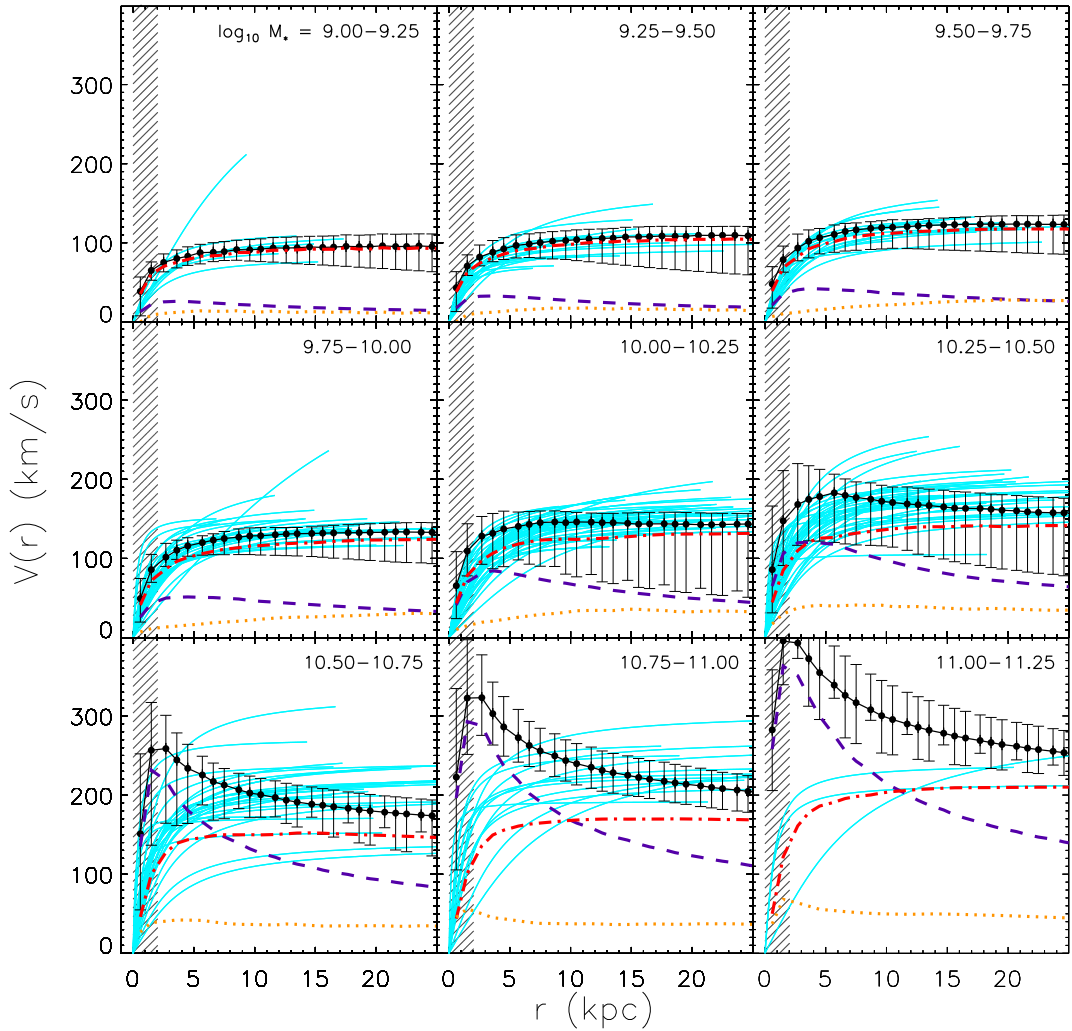}
\caption{\label{fig:vrot}
Observed rotation curves and predicted galaxy circular velocity curves in bins of galaxy stellar mass.  The continuous black curves (with filled black circles) represent the median circular velocity of the simulated galaxies and the error bars indicate the 5$^{\rm th}$ and 95$^{\rm th}$ percentiles.  The dot-dashed red, dashed purple, and dotted orange curves represent, respectively, the median circular velocity profiles of the dark matter, stars, and gas separately.  The shaded vertical region delineates the range $0 \le r \le 3$ softening lengths ($\approx 2.05$ kpc), where the circular velocities of the simulated galaxies are expected to be unreliable.
The cyan curves represent fits to the H$\alpha$ rotation curves of \citet{reyes11}.  
Over the range $9.0 \la \log_{10}[M_* (\Msun)] < 10.5$ the predicted circular velocity curves are established primarily by the dark matter and look remarkably similar to the observed rotation curves.  The simulated galaxies with $\log_{10}[M_* (\Msun)] \ga 10.6$ are rotating too rapidly in the inner regions, where the stellar component dominates the dynamics, due to overcooling (see Fig.~\ref{fig:fstar}).
}
\end{figure*}

A caveat to bear in mind when comparing the simulated and observed $V_{\rm max}-M_{*}$ relations, is that $V_{\rm max}$ for the simulated galaxies is the maximum circular velocity within $r_{200}$, whereas for the observations it is the maximum rotation velocity within the region out to which it is possible to measure gas rotation speeds (typically $\approx 20$ kpc).  We have investigated what happens to the simulated $V_{\rm max}$-$M_{*}$ relation when we limit the maximum radius to 2 effective radii (to crudely mimic the extent of the HI gas).  We find a nearly identical relationship for $\log_{10}[M_* (\Msun)] \ga 9.5$ (the measured $V_{\rm max}$ is the true one).  However, at lower masses, the radius where the circular velocity curve peaks is beyond 2 effective radii and, consequently, the measured $V_{\rm max}$ is lower than the true one, but only by $\approx -0.05$ dex on average.  This has the effect of shifting the low mass simulated galaxies down so that they tend to lie in between the results of \citet{avilareese08} (and \citealt{verheijen01}) and \citet{bell01}, yielding even better agreement between the simulations and the observations.

As discussed at the beginning of Section 3, we selected the disc-dominated systems from the simulations for comparison to the observed TF relation (of disc galaxies).  However, it is worthwhile to comment briefly on the behaviour of the spheroid-dominated galaxies, which are not shown in Fig.~\ref{fig:tf}.  Over the mass range where the disc-dominated simulated galaxies agree well with the observations ($9.0 \la \log_{10}[M_* (\Msun)] \la 10.5$), the spheroid-dominated galaxies follow the disc-dominated trend closely.  In other words, our results are insensitive to the cut in Sersic index for this mass range.  It is only at the highest masses where we see a noticeable difference, in the sense that the spheroid-dominated galaxies have lower circular velocities ($V_{80}$ and $V_{\rm max})$ at a fixed stellar mass.  However, as we will show in Section 3.3, it is clear that both disc- and spheroid-dominated galaxies at high stellar masses suffer from strong overcooling with respect to observed galaxies.

The TF relation encapsulates the rotation of a disc galaxy with a single number.  However, since neither simulated nor observed galaxies have perfectly flat rotation curves, it is possible for simulations to lie on the observed TF relation without reproducing the rotation curves of real disc galaxies.  Furthermore, it has been argued that the shapes of the rotation curves of normal disc galaxies are inconsistent with the predictions of the $\Lambda$CDM model (e.g., \citealt{salucci12}), as deduced from dark matter only simulations.
With this motivation, we next examine the rotation curves of observed and simulated galaxies as a function of galaxy stellar mass.

\subsection{Rotation curves in bins of stellar mass}

In Fig.~\ref{fig:vrot} we compare observed galaxy rotation curves with predicted galaxy circular velocity curves in 9 equal-width bins of $\log_{10} M_*$ spanning the range $9.0 \le \log_{10}[M_* (\Msun)] \le 11.25$.  The continuous black curves (connecting filled black circles) represent the median circular velocity profiles of the simulated galaxies and the error bars represent the 5$^{\rm th}$ and 95$^{\rm th}$ percentiles.  The dot-dashed red, dashed purple, and dotted orange curves represent, respectively, the median circular velocity profiles of the dark matter, stars, and gas separately.  The shaded vertical region delineates the range $0 \le r \le 3$ softening lengths ($\approx 2.05$ kpc), where the circular velocities of the simulated galaxies are expected to be unreliable.  The cyan curves represent fits to the H$\alpha$ rotation curves of \citet{reyes11}, which we reproduce using the best-fit arctan parameters provided by \citet{reyes11}.  The best-fit curves are plotted out to the radius that encloses $80\%$ of the i-band flux. 

The agreement with the amplitude and shape of the observed rotation curves over the mass range $9.0 \la \log_{10}[M_* (\Msun)] < 10.5$ is encouraging.  The observed and simulated velocity curves both show a gentle rise from $\approx 2$ kpc out to about $5$ kpc, on average, before levelling off to an approximately constant value at larger radii.  
We note that over this range in galaxy stellar mass, dark matter is the most important contributor to the circular velocity curves in the simulated galaxies (except in the innermost regions of the last mass bin), as can easily be seen by comparing the dot-dashed red curve (dark matter) to the dashed purple (stars) and dotted orange (gas) curves.

Strong deviations of the simulated circular velocity curves from the observed rotation curves are evident at higher stellar masses, however, in the sense that the simulated galaxies have mass distributions which are too concentrated, particularly in the very innermost regions ($r \la 5$ kpc).  This is clearly due to the stellar distribution being much too compact and dominant in these systems.  It is worth pointing out that many observed massive galaxies {\it do} show evidence for declining rotation velocity curves (e.g., \citealt{persic96,courteau97,verheijen01}) and that such behaviour cannot be captured by the arctan functional form adopted by \citet{reyes11} in fitting their rotation curves.  However, it is very unlikely that this accounts for most of the difference at high stellar masses.  Indeed, the fact that for these high-mass galaxies there is a normalisation offset at relatively large radii, strongly suggests that these galaxies have too high SF efficiencies.

A comparison of the rotation curve results to the TF results in Section 3.1 reveals a nice consistency, in that the rotation curves look sensible over the range of stellar mass for which the simulated galaxies lie along the TF relation, and vice versa.

In terms of the spheroid-dominated galaxies in our overall sample
(which are not used in Fig.~\ref{fig:vrot}), there are no significant
differences between disc-dominated and spheroid-dominated galaxies
in the circular velocity
curves until high stellar masses ($M_* \ga 10^{11} \Msun$) are
reached.  At high stellar masses the circular velocity curves of the
spheroid-dominated galaxies are systematically lower in amplitude and
somewhat flatter than for the disc-dominated galaxies of the same stellar mass\footnote{Naively, one might expect spheroid-dominated, low-angular momentum galaxies to have more peaked circular velocity curves than disc-dominated galaxies.  This would indeed be true for galaxies of fixed {\it stellar mass fraction}.  However, at fixed stellar mass the situation is reversed, since at the high mass end disc-dominated galaxies tend to have a higher stellar mass fraction than spheroid-dominated galaxies (see Fig.\ 4), which is found to be the case observationally (e.g., \citealt{dutton10}).  The higher stellar mass fraction leads to more peaked circular velocity curves for disc-dominated galaxies.}.

\begin{figure}
\includegraphics[width=\columnwidth]{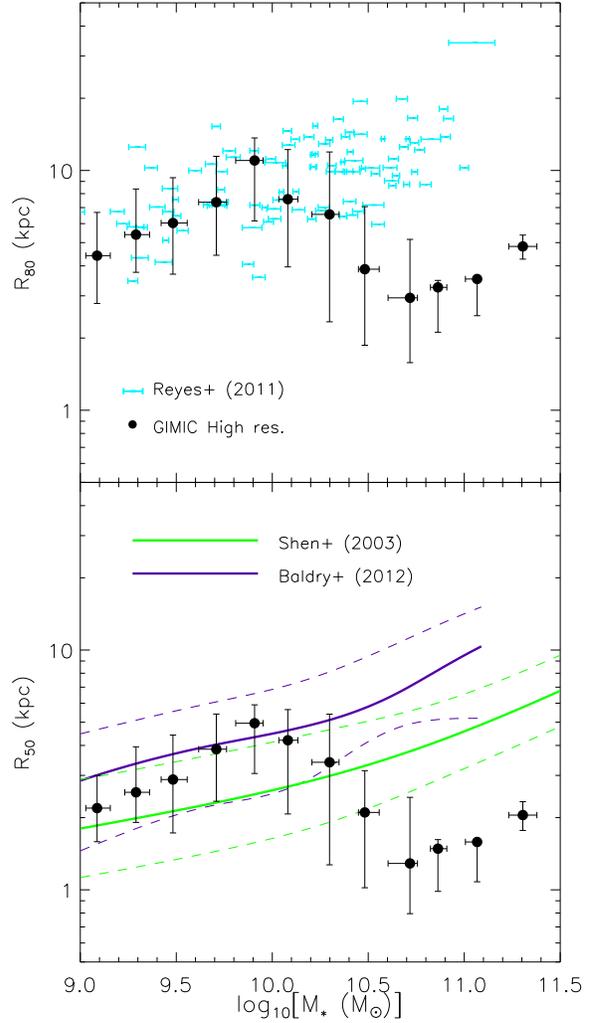}
\caption{\label{fig:size}
Comparison of simulated and observed relations between galaxy stellar mass and size, as characterised by the radii which enclose 80\% (top panel) and 50\% (bottom panel) of the i-band flux in projection.
The simulated galaxies lie approximately between the medians of the observed $M_*-R_{50}$ relations of \citet{shen03} (from SDSS data, green curves with dashed lines indicating $\pm 1\sigma$) and \citet{baldry12} (from GAMA data, purple lines, with dashed lines indicating $\pm 1\sigma$) and approximately on top of the $M_*-R_{80}$ relation of \citet{reyes11} over the range $9.0 \la \log_{10}[M_* (\Msun)] \la 10.4$.  At higher masses, the simulated galaxies are too compact compared to observed galaxies of the same stellar mass.}
\end{figure}

\subsection{Stellar mass - size relation of disc galaxies}

In Fig.~\ref{fig:size} we plot the relation between galaxy stellar mass and size.  We use the radii which enclose 80\% (top panel) and 50\% (bottom panel) of the i-band flux in projection.  The filled black circles with error bars correspond to the median and 16$^{\rm th}$ and 84$^{\rm th}$ percentiles in bins of galaxy stellar mass for the simulated \gimic\ galaxies.  The radii $R_{50}$ and $R_{80}$ are derived by fitting a Sersic law to the azimuthally-averaged i-band surface brightness profiles of the simulated galaxies, as described above.  For comparison, we plot the median $M_*-R_{50}$ {\it disc} relations (and 1-sigma scatter) of \citet{shen03} (derived from SDSS data) and \citet{baldry12} (derived from GAMA data) in the bottom panel, and individual measurements of $M_*$ and $R_{80}$ from \citet{reyes11} in the top panel.  Note that the comparison to both \citet{shen03} and \citet{baldry12} is fully consistent, in that both studies also derive $R_{50}$ by fitting Sersic laws to the azimuthally-average surface brightness profiles of the observed galaxies.  \citet{reyes11}, on the other hand, perform somewhat more complex surface brightness modelling (they fit a 2D double exponential), which we do not attempt for the simulated galaxies.

Over the range $9.0 \la \log_{10}[M_* (\Msun)] < 10.4$ the simulated galaxies lie approximately between the median relations of \citet{shen03} and \citet{baldry12} and have significant overlap with both (bottom panel).  The simulated galaxies are also consistent with data of \citet{reyes11} over this stellar mass range (taking into account error bars and scatter), although the {\it trend} with mass starts to deviate from the observations at masses of $\log_{10}[M_* (\Msun)] \sim 10.0$.  The general agreement at low-moderate masses is encouraging and, together with the agreement with the rotation curves shown above, indicates that the simulated galaxies do not suffer from the well-known `angular momentum problem'.  This is not due to resolution, as results do not change significantly if we adopt the intermediate-resolution \gimic\ simulations instead (see the Appendix).

For $\log_{10}[M_* (\Msun)] > 10.5$ the simulated galaxies are obviously too compact compared to observed galaxies of the same stellar mass.  This is consistent with our findings from the Tully-Fisher relation and the rotation curves in bins of stellar mass presented above.

\subsection{Stellar mass - star formation efficiency relation}

We have shown that over the range of stellar mass, $9.0 \la
\log_{10}[M_* (\Msun)] \la 10.5$, the \gimic\ high-resolution
simulations approximately reproduce the observed TF relation as
well as the rotation curves and galaxy sizes as a function of stellar mass.  Given that
for this mass range the rotation velocity in the simulations is
dominated by dark matter, the agreement with the observed TF relation
and rotation curves would imply that the models should have
approximately the correct SF efficiency, $\epsilon_{\rm SF} \equiv
(M_*/M_{200})/(\Omega_b/\Omega_m)$.  Here $M_*$ is the total current
stellar mass of the galaxy (which excludes surrounding satellites),
$M_{200}$ is the total mass (gas+stars+dark matter) within $r_{200}$,
and $\Omega_b/\Omega_m \approx 0.165$ is the
universal ratio of baryonic to total matter densities
\citep{komatsu11}.  The SF efficiency is therefore the fraction of the
total mass accounted for by stars, relative to the maximal case where
all the baryons are converted into stars.  Cosmological
hydrodynamic simulations have had a notoriously difficult time in
forming normal galaxies with the correct $\epsilon_{\rm SF}$,
generally generating values of 
$\epsilon_{\rm SF}$ that are a factor of
a few larger than observed (see, e.g., \citealt{guo10}, \citealt{dutton10}, and \citealt{avilareese11}, who compare the results of a number of recent hydrodynamic simulations with observations).  This is often 
referred to as the `overcooling problem' of cosmological simulations.
How do the \gimic\ high-resolution simulations fair in this regard? 

\begin{figure}
\includegraphics[width=\columnwidth]{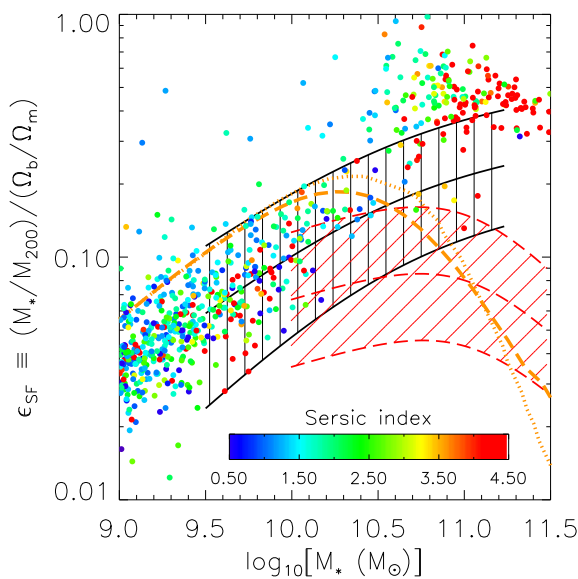}
\caption{\label{fig:fstar}
Star formation efficiency, $\epsilon_{\rm SF}$, as a function of galaxy stellar mass.  Each filled circle represents a simulated galaxy, colour-coded by the fitted Sersic index, $n$.  The dotted orange curve corresponds to the (scatter-free) relation derived by \citet{guo10} by abundance matching a stellar mass function derived from SDSS data with the halo mass function from the Millennium Simulation. The dashed orange curve is the result when a scatter of $0.15$ dex in $M_*$ at fixed $M_{200}$ is adopted in the abundance matching and $\epsilon_{\rm SF}$ is computed using the mean $M_{200}$ in bins of stellar mass. The hatched black and red regions correspond to fits of \citet{dutton10} to `direct' measures of the SF efficiency (from stacked weak lensing and satellite kinematics measurements) of disc and elliptical galaxies, respectively (see text).  The simulated disc galaxies match the observed $\epsilon_{\rm SF}$ of disc galaxies over the mass range $9.0 \le \log_{10}[M_* (\Msun)] < 10.5$.  At higher masses, the simulated elliptical galaxies have too high $\epsilon_{\rm SF}$ relative to what is inferred observationally.
}
\end{figure}

In Fig.~\ref{fig:fstar} we plot the trend between $\epsilon_{\rm SF}$ and stellar mass for the simulated galaxies at $z=0$.  
Each of the dots in Fig.~\ref{fig:fstar} represents a simulated galaxy.  The dots are colour-coded according to the fitted Sersic index.  Note that we include both disc- and spheroid-dominated galaxies in this plot.
The dotted orange curve is the relation derived by \citet{guo10} by applying the abundance matching technique to the SDSS galaxy stellar mass function\footnote{The stellar masses of \citet{li09} were derived using the five-band SDSS photometric data (and redshift) as described in \citet{blanton07} and assuming a Chabrier IMF, consistent with what is adopted in our cosmological simulations.} derived by \citet{li09} and the dark matter halo mass function from the Millennium Simulation.  This result assumes no scatter in the relation between $M_*$ and $M_{200}$.  The dashed orange curve is the result when a scatter of $0.15$ dex in $M_*$ at fixed $M_{200}$ is adopted and $\epsilon_{\rm SF}$ is computed using the mean $M_{200}$ in bins of stellar mass (Guo, priv.\ comm.).  The shaded black and red regions correspond to the (by eye) fit of \citet{dutton10} to `direct' measurements\footnote{We note the `direct' measurements of halo mass are model-dependent, in that they assume a NFW halo profile with a particular mass-concentration relation that does not account for possible correlations between the scatter in concentration and galaxy morphology/colour.} of the SF efficiencies of disc and elliptical galaxies, respectively.  The direct measurements are based on stacked weak lensing \citep{mandelbaum06,mandelbaum08,schulz10} and stacked satellite kinematics \citep{conroy07,klypin09,more11} (hereafter WL/SK) analyses of disc and elliptical galaxies.  
By stacking the WL/SK of galaxies of fixed stellar mass (or luminosity), these authors derive the mean halo mass (e.g., $M_{200}$) as a function of stellar mass and thereby constrain the SF efficiency.  Importantly, \citet{dutton10} have taken care to adjust the stellar masses from these studies so that they all correspond to the Chabrier IMF, and to ensure that the halo mass is defined as $M_{200}$ throughout (see \citealt{dutton10} for details).  Note also that the stellar masses derived in the individual WL/SK studies are typically based on the single colour scalings of \citet{bell03} or \citet{bell01}.  We therefore apply a small mass-dependent adjustment so that stellar masses from these studies are consistent with those derived from five-band SDSS data \citep{blanton07} (as described in Section 1).  
It is therefore possible to directly compare the WL/SK estimates of the SF efficiency (the shaded regions) with the abundance matching results represented by the dashed and dotted black curves.  The shaded regions of \citet{dutton10} bracket the 2-sigma (95\% CL) errors quoted in the individual WL/SK studies.

Consistent with the results derived in Sections 3.1 and 3.2, the simulated galaxies have SF efficiencies that are quite comparable to what is derived from stacked WL/SK analyses of disc galaxies over the range in galaxy mass, $9.0 \la \log_{10}[M_* (\Msun)] < 10.5$.  The correct fraction is achieved through efficient (but energetically feasible) SN feedback which ejects gas from the system (see, e.g., Fig.\ 7 of \citealt{crain10}).  This is ultimately what also allows the simulations to reproduce the TF relation and the shape and normalization of the rotation curves over this mass range.  We point out that unlike some recent studies of individual galaxies (e.g., \citealt{stinson10,maccio12,brook12}), we have not had to invoke multiple sources of feedback in order to reproduce the observed SF efficiency.  All that has been done is to adjust the mass-loading of the SN feedback so that the simulations roughly reproduce the peak of the cosmic SF rate density of the Universe.

Interestingly, most galaxies in the simulations fall {\emph{below}} the
relation derived from applying the abundance matching technique to the
galaxy stellar mass function (with or without scatter included) for
$M_* \la 10^{10} \Msun$.  \citet{dutton10} previously pointed out
the factor of $\approx 2$ inconsistency between the abundance matching
and direct measure studies.  It is presently unclear what the origin
of the discrepancy is.  The discrepancy cannot be explained by
differences in the cosmological parameters adopted for the Millennium
Simulation (from which \citealt{guo10} derive their halo mass
function) and current best estimates, as \citet{dutton10} have shown
that the discrepancy exists for other abundance matching-based results
(e.g., \citealt{moster10,behroozi10}) which adopted parameter values
more in line with those derived from recent CMB analyses.
Furthermore, the discrepancy is not simply due to the fact that the abundance matching
technique is applied to the stellar mass function without regard to 
galaxy morphology, since disc galaxies dominate the stellar mass
function at stellar masses where the discrepancy exists (thus, the
abundance matching results should reflect those of disc galaxies at
these masses). 

However, the fact that the simulations approximately match the TF relation and velocity rotation curves over the same range of stellar mass, suggests that the systematic offset may lie mostly with the abundance matching results.  Consistent with this hypothesis is the normalisation offset between the TF relation deduced from the abundance matching technique\footnote{\citet{guo10} use abundance matching to derive a relation between $M_*$ and $M_{200}$ and use the relation between $V_{\rm max}$ and $M_{200}$ from the Millennium Simulation to derive the relation between $M_*$ and $V_{\rm max}$.  If dark matter dominates the rotation curves, then this is equivalent to the TF relation.} and the observed relation (see \citealt{guo10}).  \citet{guo10} suggested that the offset could be due to the fact that their calculations did not take into account adiabatic contraction of the dark matter halo due to centrally-concentrated cold baryons, but an alternative explanation is that the dark matter has not significantly contracted (see, e.g., \citealt{duffy10,dutton11,maccio12} and Section 3.5 below) and that the abundance matching results systematically overestimate the stellar masses by a factor of $\approx 2$.

Indeed, the agreement with the TF relation in the top panel of
Fig.~\ref{fig:tf} implies that the simulated galaxies have the correct
stellar mass fractions within $r_{80}$ over the range $9.0 \la \log_{10}[M_*
(\Msun)] < 10.5$, since $V_{80}$ is a direct measure of the total mass (stars+gas+DM) within that radius (under the assumption that non-circular motions are small).  Assuming that the observed and simulated galaxies have similar (dark) matter distributions beyond this radius (which is implied by the WL/SK data since these shows that an NFW profile \citep{nfw96,nfw97} works well at these radii), there is therefore consistency between the match to the WL/SK data in Fig.~\ref{fig:fstar} and the TF relation in the top panel of Fig.~\ref{fig:tf}.

A possible systematic error in the abundance matching formalism is the implicit assumption that satellite and central galaxies of fixed halo mass have the same\footnote{Note, however, that usually abundance matching is performed for satellite galaxies at the time of accretion.} stellar mass.  Indeed, \citet{rod11} have shown that by excluding satellites from the abundance matching calculation (i.e., both from the halo mass function and the observed stellar mass function) the star formation efficiency of central galaxies decreases slightly, which goes in the right direction to resolve the difference between the abundance matching results and those based on WL/SK.  However, the effect appears to be too small (only $\approx10-20\%$) to reconcile the difference.

There is a small number of apparent outliers from the main $\epsilon_{\rm SF} - M_*$ trend at low to moderate stellar masses.  We have investigated the nature of these outliers and found that they always correspond to galaxies in close proximity to a massive galaxy group.  The dark matter associated with the galaxy is clearly truncated (which is why $\epsilon_{\rm SF}$ is higher than typical).  This may be because these galaxies have passed through the more massive neighbour (and hence have been tidally stripped) in the past, but have since come back out to larger radii (see the population of ``backsplash'' subhaloes on extreme orbits identified in cosmological simulations by \citealt{gill05,ludlow09}).

Returning to the general trend, for stellar masses $\log_{10}[M_* (\Msun)] \ga 10.5$ the simulations clearly have too high stellar mass fractions with respect to both the abundance matching results and the stacked measurements of disc and especially elliptical galaxies.  Furthermore, the simulations do not produce a bi-modal distribution of stellar mass fractions (corresponding to discs and ellipticals) over the range $10.5 \la \log_{10}[M_* (\Msun)] \le 11.0$ as implied by stacked WL/SK measurements.  It is conceivable that feedback from AGN, which was not included in our simulations and which is expected to become relatively more important at high masses (and for galaxies with relatively large spheroidal components), may resolve these issues.  Indeed, \citet{mccarthy10} have used simulations from the OverWhelmingly Large Simulations project \citep{schaye10}, which uses the same simulation code as \gimic, to show that the inclusion of AGN feedback in cosmological simulations can reduce the stellar mass of the most massive galaxies by up to an order of magnitude (see Fig.\ 7 of that study), yielding agreement with observations on the mass scales corresponding to groups.  Finally, we point out that current state-of-the-art very high resolution zoomed simulations (that neglect AGN feedback) also suffer from overcooling at these high masses\footnote{The code used for the \gimic\ simulations was also used in the Aquila project, which simulated a single $M_{200} = 1.6 \times 10^{12} \Msun$ halo.  The results of that individual galaxy are completely in line with the trends present in Figs.\ 1 and 3 of the present study.}, as demonstrated by the Aquila project \citep{scannapieco11}.

\begin{figure}
\includegraphics[width=\columnwidth]{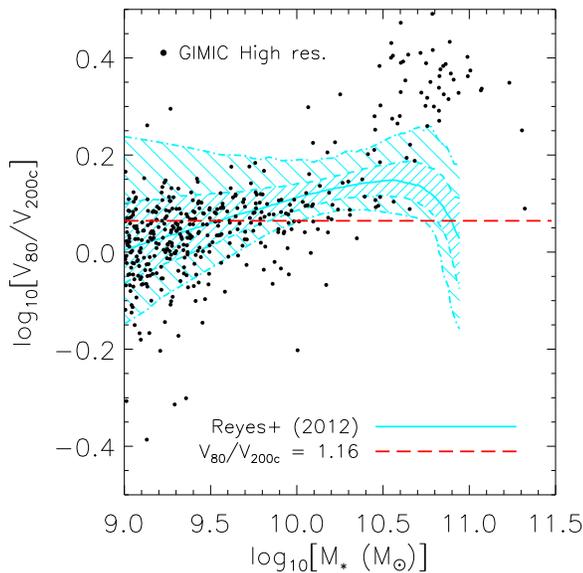}
\caption{\label{fig:vratio}
Comparison of simulated and observed relations between galaxy stellar mass and the 
ratio of the `optical rotation velocity' (defined here as $V_{80}$) to virial circular velocity, $V_{200c} \equiv \sqrt{G M_{200} / r_{200}}$, for disc galaxies at $z=0$.  The shaded regions encompass the $\pm 1$ sigma (fine shading) and $\pm 2$ sigma (spare shading) distributions of \citet{reyes12}.
The simulated galaxies lie approximately on top of the observed relation over the wide range $9.0 \la \log_{10}[M_* (\Msun)] \la 10.5$.  At higher masses, the simulated galaxies have circular velocities that are too large with respect to their virial circular velocities.
}
\end{figure}

\subsection{Stellar mass - $V_{80}/V_{200}$ relation}

Related to the star formation efficiency, $\epsilon_{\rm SF}$, is the ratio of galaxy rotation velocity to its virial circular velocity, which we define here (following \citealt{reyes12}; see also \citealt{dutton10}) as $V_{80}/V_{200c}$, where $V_{200c} \equiv \sqrt{G M_{200} / r_{200}}$.  If baryonic processes do not significantly affect the total matter distribution at small radii (i.e., so that it approximately traces an unmodified NFW profile) we would expect\footnote{Here we used the $z=0$ Millennium Simulation halo catalog to measure the median value of $V_{\rm max}/V_{200c}$ (1.16) for haloes in the mass range $11.0 \le \log_{10}[M_{200} (\Msun)] \la 13.0$.  Note that the ratio $V_{\rm max}/V_{200c}$ varies with halo mass, owing to the fact that lower mass haloes are more concentrated on average, but the dependence is very weak.} $V_{80}/V_{200c} \approx V_{\rm max}/V_{200c} \approx 1.16$ typically for this range of masses.  If, on the other hand,  baryons significantly contract (expand) the potential well on scales similar to the optical extent of the galaxy, this will have the effect of raising (lowering) the $V_{\rm opt}/V_{\rm vir}$ (e.g., \citealt{duffy10}).

In Fig.~\ref{fig:vratio} we examine the relation between galaxy stellar mass and $V_{\rm opt}/V_{\rm vir} = V_{80}/V_{200c}$ for our simulated disc galaxies.  We make comparisons to the observations of \citet{reyes12}, who measured this relation for a disc galaxy sample that is representative of the Tully-Fisher sample of \citet{reyes11} (see Fig.~\ref{fig:tf}).  \citet{reyes12} employed stacked galaxy-galaxy lensing to measure a mean $M_{200}$ (and thus a mean $V_{200c}$) in bins of stellar mass.  Consistent with our previous results, the simulated galaxies lie approximately on top of the observed relation for $9.0 \la \log_{10}[M_* (\Msun)] \la 10.5$, while at higher masses the simulated galaxies have too high galaxy circular velocities with respect to the virial circular velocity of the halo in which they live.

The red dashed line in Fig.~\ref{fig:vratio} represents the ratio expected in the absence of significant contraction or expansion of the inner potential well.  If we restrict the comparison to the mass range $9.0 \la \log_{10}[M_* (\Msun)] \la 10.5$, both the \gimic\ simulations and the observations are consistent with a mild degree of expansion at low stellar masses ($\log_{10}[M_* (\Msun)] \la 9.5$) and a similarly mild amount of contraction at higher stellar masses ($\log_{10}[M_* (\Msun)] \ga 10.0$).  At very high masses ($\log_{10}[M_* (\Msun)] \ga 10.5$), the simulated potential wells have clearly undergone too much contraction as a result of overcooling.

\subsection{Stellar mass - sSFR relation}

As discussed in the Introduction, there has been considerable progress in recent years in the ability of simulations to reproduce the structural and dynamical properties of normal disc galaxies at $z=0$.  This is due in large part to the implementation of strong feedback mechanisms in the simulations.  However, it has been pointed out that in spite of this progress, there remain a number of important problems, including the apparent inability of simulations with strong feedback to reproduce the sSFRs of galaxies locally and at high redshift, particularly for low-mass galaxies (e.g., \citealt{dave11,avilareese11}; but see \citealt{brook12}).  Here we examine the trend between galaxy stellar mass and sSFR at $z=0$ for disc galaxies in the \gimic\ simulations.

\begin{figure}
\includegraphics[width=\columnwidth]{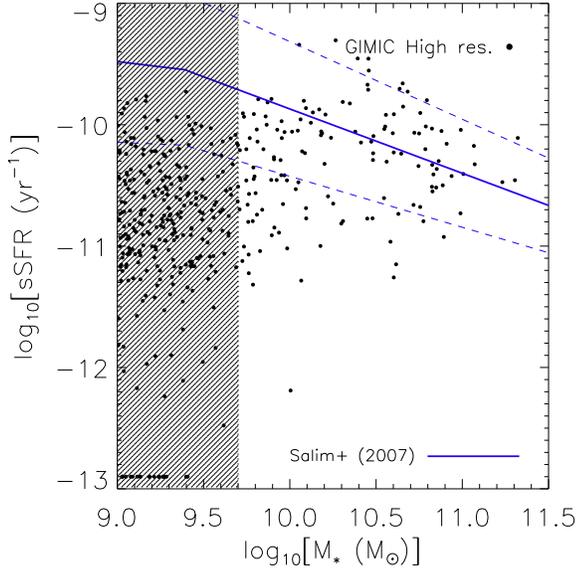}
\caption{\label{fig:ssfr} Comparison of simulated and observed relations between stellar mass and the sSFR of disc galaxies at $z=0$.  The blue solid and dashed lines represent the median and $\pm 1$ sigma intrinsic scatter, respectively, of the observed `blue galaxy' sample of \citet{salim07}.  The black shaded region represents the galaxy mass range for which sSFRs are sensitive to resolution (i.e., are likely not converged - see the Appendix).  For the simulated disc galaxies that are not currently forming stars (i.e., SFR$=0$), we arbitrarily set their $\log_{10}$[sSFR]$=-12.9$ in this plot.   There is rough agreement with the observed sSFRs for simulated galaxies with $\log_{10}[M_* (\Msun)] \ga 10.0$.
}
\end{figure}

In Fig.~\ref{fig:ssfr} we compare the relation between galaxy stellar mass and sSFR for disc galaxies in the \gimic\ simulations with the observations of \citet{salim07}.  \citet{salim07} estimated star formation rates using UV data from {\it GALEX} for a large sample of $\sim$ 50,000 optically-selected galaxies from the SDSS.  The black shaded region in Fig.~\ref{fig:ssfr} represents the galaxy mass range for which the sSFRs are sensitive to numerical resolution and are likely not converged.  

At stellar masses of $\log_{10}[M_* (\Msun)] \ga 10.0$, the simulated galaxies roughly reproduce the observed trend, which is encouraging.  Note that this (rough) agreement was not a given, as the feedback parameters in \gimic\ have been tuned to match only the peak of the cosmic star formation rate density (at $z \sim 1-3$), not the star formation rates of galaxies at the present day.  Note, however, that the agreement at the highest masses ($\log_{10}[M_* (\Msun)] \ga 10.5$) must be fortuitous, as we have already demonstrated that these galaxies suffer from overcooling (therefore the star formation rates of these galaxies are too high given their halo masses).

At lower stellar masses ($\log_{10}[M_* (\Msun)] \la 10.0$) the simulated galaxies fall below the observed relation of \citet{salim07}.  However, as we demonstrate in the Appendix, the sSFRs of galaxies with such low masses are rather poorly converged at $z=0$ in our simulations.
%
%
It is possible that a further increase in resolution would result in yet higher sSFRs, possibly resolving the discrepancy at low stellar masses. Consistent with this hypothesis is the fact that \citet{brook12} have used very high resolution zoomed simulations of low-mass galaxies and are able to reproduce the relation
of \citet{salim07} down to low stellar masses. Thus, it remains to be seen whether the problem of some simulations in not reproducing the sSFRs of low-mass galaxies is a fundamental one.

\section{Connecting star formation efficiency, the stellar mass function, and the TF relation}

In this section we briefly comment on the link between the star formation efficiency (i.e., the $M_*-M_*/M_{200}$ relation), the galaxy stellar mass function, and the TF relation.  

The virial mass of a halo of mass $M_{200}$ can be expressed in terms of its virial circular velocity, $V_{200} \equiv (G M_{200} / r_{200})^{1/2}$, as

\begin{equation}
M_{200} = 3.2 \times 10^{11} \Msun \biggl(\frac{V_{200}}{100 \ {\rm km s}^{-1}}\biggl)^3 \,
\end{equation}

\noindent assuming a Hubble constant of $H_0 = 72$ km s$^{-1}$ Mpc$^{-1}$ (e.g., \citealt{mo98}).  Thus, if we knew the relation between galaxy stellar mass, $M_*$, and $M_{200}$ we could convert eqn.\ (2) into a TF relation.

For halos with masses below $M_{200} \sim 10^{14} \Msun$, the halo mass function is well represented by a power-law of the form 

\begin{equation}
\frac{dn}{dM_{200}} \propto M_{200}^{-1.9} 
\end{equation}

\noindent (e.g., \citealt{jenkins01,reed07}). If we assume the galaxy stellar mass function also follows a power-law of the form 
\begin{equation}
\frac{dn}{dM_*} \propto M_*^{-\alpha} \ ,
\end{equation}

\noindent then matching abundances of haloes and galaxies\footnote{Note that in Section 3.4 we argued there may be a systematic problem with galaxy stellar masses inferred through the abundance matching technique.  However, this appears to be a problem in the normalisation, rather than in the slope (which is very similar to that derived from stacked WL/SK).  Here we are interested in how the slopes of the $\epsilon-M_*$ relation, the TF relation, and the stellar mass function are related.} results in an $\epsilon_{\rm SF}$-$M_{200}$ relation of
\begin{equation}
\epsilon_{\rm SF} \equiv \frac{M_*}{M_{200}} \propto M_*^{[(1.9-\alpha)/0.9]}  \ . 
\end{equation}
 
\noindent Inserting the above into eqn.\ (2) yields a TF relation of the form
\begin{equation}
V_{200} \propto M_*^{(1/3)(\alpha-1)/0.9} \propto M_*^{(\alpha-1)/2.7} \ \ \ ,
\end{equation}

Over the range $9.0 < \log_{10}[M_* (\Msun)] < 10.0$ (which avoids resolution issues at lower masses and the knee in the stellar mass function at higher masses, which would violate the power-law assumption in eqn.\ 4), the galaxy stellar mass function in \gimic\ is characterised relatively well by a power-law with $\alpha \approx 1.6$.  From eqns.\ (5) and (6) this yields $\epsilon_{\rm SF} \propto M_*^{\approx 1/3}$ and $V_{200} \propto M_*^{\approx 0.22}$ over the same range in stellar mass.  These expectations match what we actually measure over this range in \gimic\ well.

It is therefore possible, at least in principle, to constrain the faint-end slope of the galaxy stellar mass function from the slope of the TF relation (or vice versa).  On this basis, the observed TF relations in Fig.\ 1 suggest that $\alpha$ must be relatively steep, $\ga 1.5$, which is close the value of $1.45$ found by \citet{baldry12} for blue galaxies.  

A rigorous statement to this effect would necessitate relaxing a number of assumptions.  Firstly, we have assumed power-law dependencies for the halo and stellar mass functions over this mass range.  This holds to a high level of accuracy for the halo mass function but should be checked for the stellar mass function.  Secondly, we have assumed that there is no scatter in the $M_*$-$M_{200}$ relation.  We have explored the effects of adding log-normal scatter to an idealised powerlaw relation between $M_*$ and $M_{200}$ and find that it does not bias the recovered slope of the resulting stellar mass function.  However, if the amplitude of the scatter were a strong function of $M_{200}$ and/or if the scatter were not log-normal, then this can result in a bias in $\alpha$.  Finally, we have derived the TF relation in eqn.\ 6 in terms of the virial circular velocity, whereas what is measured is closer to $V_{\rm max}$.  If dark matter dominates the dynamics, then to a good approximation $V_{\rm max} \approx 1.2 V_{200}$.  This breaks down when applied to a very wide range of halo masses owing to the mass-concentration relation of dark matter haloes, an effect, however, which is unimportant for the mass range under consideration here.  Potentially more relevant here is that baryonic physics can either contract or expand the dark matter distribution (e.g., \citealt{duffy10,dutton11}) and can therefore introduce a halo mass dependence in the relationship between $V_{\rm max}$ and $V_{200}$.  In the \gimic\ simulations, we find these effects to be small over the mass range considered here (i.e., $9.0 < \log_{10}[M_* (\Msun)] < 10.0$).

\section{Summary and Discussion}

We have used the \gimic\ simulations to study the mass distribution and SF efficiency of a large, cosmologically representative population of simulated disc galaxies over the wide range of stellar mass $9.0 \le \log_{10}[M_* (\Msun)] \le 11.5$ at $z=0$.  These simulations include efficient, but energetically feasible supernova feedback, but have not been tuned in any way to produce ``realistic'' disc galaxies.  The main results of our study may be summarised as follows:

\begin{itemize}
\item{Over the stellar mass range $9.0 < \log_{10}[M_* (\Msun)] <
    10.5$, the simulations approximately reproduce: the observed Tully-Fisher
    relation (this is true for both the velocity measured at the
    radius that encloses 80\% of the galaxy's i-band light, $V_{80}(M_*)$, 
    and the peak velocity, $V_{\rm max}(M_*)$), the observed rotation
    velocity curves of disc galaxies, the observed size-mass relation of disc galaxies, the relation stellar mass ($M_*$) and optical rotation to virial circular velocity ratio (`$V_{\rm opt}/V_{\rm vir}$'), and the trend between SF
    efficiency and galaxy stellar mass, as inferred from stacked weak
    lensing and stacked satellite kinematics studies.  Over this mass
    range, dark matter dominates the rotation curves of galaxies.}
\item{A comparison of both the observed and simulated values of `$V_{\rm opt}/V_{\rm vir}$' with that predicted by dark matter only simulations indicates that the potential wells of galaxies with relatively low stellar masses ($\log_{10}[M_* (\Msun)] \la 9.5$) have undergone a relatively mild amount of expansion, while the potential wells of galaxies with higher stellar masses ($\log_{10}[M_* (\Msun)] \ga 10.0$) have undergone a similarly mild degree of contraction (see Sec. 3.6).}
\item{For stellar masses $\log_{10}[M_* (\Msun)] \ga 10.6$, the
    simulated galaxies rotate faster than observed
    galaxies of the same stellar mass, are too compact, and have too large a fraction
    of their baryons locked up in stars (i.e., the simulations suffer
    from overcooling).}
\item{The simulations also reproduce the specific star formation rates of $\sim L*$ galaxies (with $\log_{10}[M_* (\Msun)] \sim 10.6$) but have too low levels of SF in lower mass galaxies, which is plausibly due to the finite resolution of the simulations.}
 \item{There is an intriguing discrepancy between the SF efficiency of disc
    galaxies derived from abundance matching and `direct' measurements
    derived from stacked weak lensing/satellite kinematics for $M_*
    \la 10^{10} \Msun$, which was previously pointed out by
    \citet{dutton10}.  The fact that our simulations simultaneously
    reproduce the TF relation, rotation curves in bins of stellar
    mass, and the SF efficiency derived from direct measurements over
    this mass range, suggests that the unknown systematic may lie
    mostly with the abundance matching results.  One possibility is that the implicit assumption that satellites (at accretion) and central galaxies of fixed halo mass have the same star formation efficiency is incorrect (e.g., \citealt{rod11}.}

\item{Simple scaling arguments presented in Section 4 (see also \citealt{mo10}) show that it is possible to constrain the faint-end slope of the galaxy stellar mass function from the slope of the TF relation (or vice versa).  Current measurements of the TF relation suggest that the faint slope should be relatively steep ($\alpha \ga 1.5$), which is consistent with some recent measurements (e.g., \citealt{baldry12}).} 
\end{itemize}

The success of the \gimic\ simulations in reproducing the observed trends at low to moderate stellar masses is directly attributable to efficient SN feedback.  As shown previously by \citet{crain10}, the SN feedback in \gimic\ drives vigorous galactic winds, particularly at high redshift, which eject a significant fraction of the baryons from haloes with masses $\log_{10}[M_{200} (\Msun)] \la 12.5$ and pollute the intergalactic medium with metals.  Thus, comparisons with observations of galactic outflows and the intergalactic medium represent promising further ways to test the model.  We point out that efficient feedback was achieved without the need for extremely high resolution and is a result of the way in which the feedback was implemented in the simulations.  Recent zoomed simulations in which the feedback was injected thermally and the cooling turned off temporarily (e.g., \citealt{governato10,stinson10,guedes11,brook12,maccio12}), appear to require much higher resolution before the feedback becomes similarly efficient.

For $\log_{10}[M_* (\Msun)] \ga 10.6$, the galaxies in the \gimic\ simulations suffer from
strong overcooling, as do all other current cosmological simulations
of disc galaxy formation that invoke only SN feedback.  We conjecture
that another energetically important form of feedback is required at
high masses, the likely candidate being that from accreting
supermassive black holes \citep{benson03,dimatteo05,bower06,croton06,booth09}.
\citet{mccarthy10} have shown, using simulations that form part of the
OverWhelmingly Large Simulations project \citep{schaye10}, that AGN
feedback is crucial in reproducing the hot gas and stellar component
of systems with halo masses only slightly higher than explored here
(corresponding to galaxy groups).  It is therefore certainly plausible
that AGN are relevant over the mass scale where \gimic\ (and other
current simulations) fails.

While the results of the present study certainly give reason for optimism, any successful model must not only reproduce these (and other) trends at $z=0$ but also their evolution to high redshift, which is by no means automatic.  In a future study we intend to explore a wider range of observables (for both discs and ellipticals) over a wider range of redshifts and with a wider range of models, some of which include feedback from AGN.

\section*{Acknowledgements}

The authors thank the referee, Vladimir Avila-Reese, for a helpful and constructive report.  They also thank Qi Guo for providing her abundance matching results, Ivan Baldry for providing the GAMA stellar mass-size relation, and Aaron Dutton, Frank van den Bosch, Reina Reyes, Chris Brook, and Fabio Governato for helpful comments.  IGM is supported by an STFC
Advanced Fellowship at the University of Birmingham.  The simulations
presented here were carried out by the Virgo Consortium for
Cosmological Supercomputer Simulations using the HPCx facility at the
Edinburgh Parallel Computing Centre (EPCC) as part of the EC's DEISA
`Extreme Computing Initiative', the Cosmology Machine at the Institute
for Computational Cosmology of Durham University, and Darwin at the
University of Cambridge.  CSF acknowledges a Royal Society Wolfson
Research Merit Award. This work was supported by Marie Curie Initial
training Network CosmoComp (PITN-GA-2009-238536), ERC Advanced
Investigator grant COSMIWAY and an STFC rolling grant to the Institute
for Computational Cosmology.  This research was also supported in part by the National Science Foundation under Grant No. NSF PHY11-2595.

\begin{figure}
\includegraphics[width=\columnwidth]{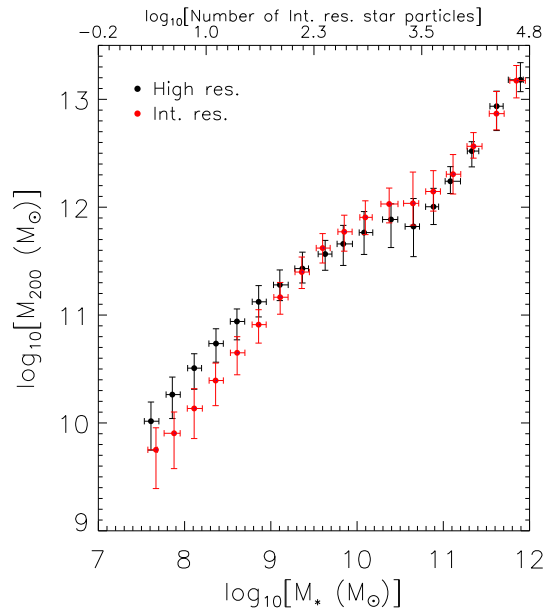}
\caption{\label{fig:res1}
Comparison of the $M_{*}-M_{200}$ relations for the \gimic\ intermediate- and high-resolution simulations.  The filled circles correspond to the median halo mass, $M_{200}$, in bins of galaxy stellar mass, $M_*$, while the error bars correspond to the 14th and 86th percentiles.  There is reasonably good agreement when there are a minimum of $\sim 100$ star particles.  We conservatively impose a cut of $500$ star particles on the high-resolution simulation in the analyses presented in this paper.
}
\end{figure}

\begin{figure}
\includegraphics[width=\columnwidth]{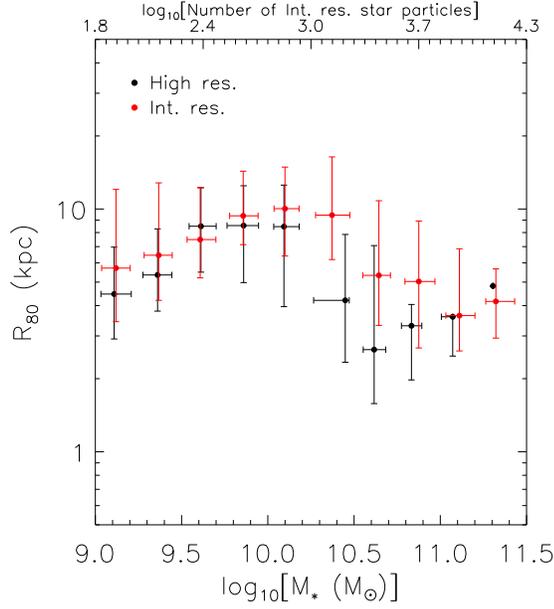}
\caption{\label{fig:res2}
Comparison of the $M_{*}-R_{80}$ relations of disc-dominated galaxies for the \gimic\ intermediate- and high-resolution simulations.  The filled circles correspond to the median value of $R_{80}$ in bins of galaxy stellar mass, $M_*$, while the error bars correspond to the 14th and 86th percentiles.  There is reasonably good agreement over the mass range considered here.
}
\end{figure}

\begin{figure}
\includegraphics[width=\columnwidth]{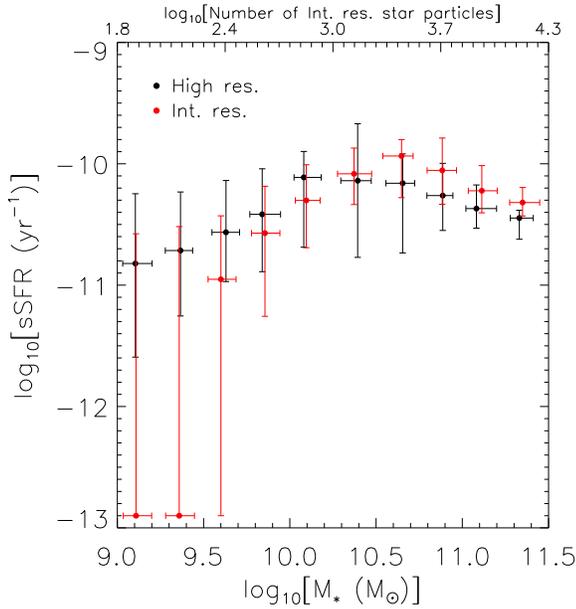}
\caption{\label{fig:res3}
Comparison of the $M_{*}-$sSFR relations of disc-dominated galaxies for the \gimic\ intermediate- and high-resolution simulations.  The filled circles correspond to the median sSFR in bins of galaxy stellar mass, $M_*$, while the error bars correspond to the 14th and 86th percentiles.  There is reasonably good agreement for galaxies with $\log_{10}[M_* (\Msun)] \ga 9.7$.  At lower masses, the sSFR rate decreases significantly with decreasing resolution. 
}
\end{figure}

\section*{Appendix: Numerical convergence}

Here we investigate the sensitivity of our results to the numerical resolution.  In Fig.~\ref{fig:res1} we compare the $M_{*}-M_{200}$ relations of the \gimic\ intermediate- and high-resolution simulations.  In general there is good agreement when there is a minimum of $\sim 100$ star particles (modulo some sensitivity to the exact mass scale where feedback stalls).  We conservatively impose a cut of $500$ star particles on the high-resolution simulation (corresponding to $\log_{10}[M_* (\Msun)] = 9.0$) in the paper.

In Fig.~\ref{fig:res2} we compare the $M_{*}-R_{80}$ relations of the \gimic\ intermediate- and high-resolution simulations.  Again, there is reasonably good agreement when there is a minimum of $\sim 100$ star particles, but with some differences occurring at the transition from the effective to ineffective feedback regimes.

In Fig.~\ref{fig:res3} we compare the $M_{*}-$sSFR relations of the \gimic\ intermediate- and high-resolution simulations.  There is reasonably good agreement for galaxies with $\log_{10}[M_* (\Msun)] \ga 9.7$.  At lower masses, the sSFR rate decreases significantly with decreasing resolution.

\end{document}